\documentclass{Interspeech}

\interspeechcameraready

\title{Pushing the Frontiers of Self-Distillation Prototypes Network with Dimension Regularization and Score Normalization}

\author[affiliation={1}]{Yafeng}{Chen}
\author[affiliation={1}]{Chong}{Deng}
\author[affiliation={1}]{Hui}{Wang}
\author[affiliation={1}]{Yiheng}{Jiang}
\author[affiliation={1}]{Han}{Yin}
\author[affiliation={1}]{Qian}{Chen}
\author[affiliation={1}]{Wen}{Wang}

\affiliation{Tongyi Speech Lab}{Alibaba Group}{China}
\email{chenyafeng.cyf@alibaba-inc.com}
\keywords{speaker verification, self-distillation prototypes network, dimension regularization, score normalization}

\usepackage{comment}
\usepackage{cite}
\usepackage{amsmath,amssymb,booktabs}
\usepackage{multirow}

\begin{document}

\maketitle

\begin{abstract}
    Developing robust speaker verification (SV) systems without speaker labels has been a longstanding challenge. Earlier research has highlighted a considerable performance gap between self-supervised and fully supervised approaches. In this paper, we enhance the non-contrastive self-supervised framework, Self-Distillation Prototypes Network (SDPN), by introducing dimension regularization that explicitly addresses the collapse problem through the application of regularization terms to speaker embeddings. Moreover, we integrate score normalization techniques from fully supervised SV to further bridge the gap toward supervised verification performance. SDPN with dimension regularization and score normalization sets a new state-of-the-art on the VoxCeleb1 speaker verification evaluation benchmark, achieving Equal Error Rate \textbf{1.29\%}, \textbf{1.60\%}, and \textbf{2.80\%} for trial VoxCeleb1-\{O,E,H\} respectively.\footnote{Code will be publicly available at \url{https://github.com/modelscope/3D-Speaker}} These results demonstrate relative improvements of \textbf{28.3\%}, \textbf{19.6\%}, and \textbf{22.6\%} over the current best self-supervised methods, thereby advancing the frontiers of SV technology.
\end{abstract}

\vspace{-0.1cm}
\section{Introduction}

Deep learning methods have significantly advanced the performance of speaker verification (SV) tasks. These improvements have been driven by the availability of large labeled datasets and effective data augmentation methods. However, collecting extensive labeled data in the real world is both time-consuming and costly. As an alternative, self-supervised learning (SSL), which relies on unlabeled data, offers a promising solution for learning robust speech representations.  Self-supervised methods are typically categorized into three main directions: contrastive~\cite{DBLP:conf/icml/ChenK0H20, DBLP:conf/cvpr/He0WXG20, DBLP:conf/icassp/XiaZWYY21, DBLP:conf/icassp/ZhangJCLHS22, DBLP:conf/nips/0004LW023, DBLP:journals/taslp/TuMC24}, non-contrastive~\cite{DBLP:conf/nips/GrillSATRBDPGAP20,DBLP:conf/iccv/CaronTMJMBJ21,DBLP:conf/icassp/SangLLAW22, DBLP:journals/jstsp/ZhangY22, DBLP:conf/icassp/ChenZWCC23, DBLP:conf/icassp/Chen25}, and dimension contrastive~\cite{DBLP:conf/icml/ZbontarJMLD21,DBLP:conf/iclr/BardesPL22, DBLP:conf/iclr/GarridoCBNL23} learning.

Contrastive learning frameworks~\cite{DBLP:conf/icml/ChenK0H20, DBLP:conf/cvpr/He0WXG20, DBLP:conf/icassp/XiaZWYY21, DBLP:conf/icassp/ZhangJCLHS22, DBLP:conf/nips/0004LW023, DBLP:journals/taslp/TuMC24}, such as SimCLR~\cite{DBLP:conf/icml/ChenK0H20} and MoCo~\cite{DBLP:conf/cvpr/He0WXG20}, establish instance-level discriminative capabilities through positive/negative pair optimization. While effective in learning representations that are invariant to data augmentation, these methods face challenges due to their reliance on large batch sizes (SimCLR) or complex memory mechanisms (MoCo), which lead to computational bottlenecks. Xia et al.\cite{DBLP:conf/icassp/XiaZWYY21} introduce SimCLR and MoCo frameworks with designed augmentation strategies for self-supervised speaker verification, achieving performance improvements. 

Non-contrastive methods~\cite{DBLP:conf/nips/GrillSATRBDPGAP20,DBLP:conf/iccv/CaronTMJMBJ21,DBLP:conf/icassp/SangLLAW22, DBLP:journals/jstsp/ZhangY22, DBLP:conf/icassp/ChenZWCC23, DBLP:conf/icassp/Chen25} avoid negative pairs,
introducing the collapse problem where speaker representations converge to trivial solutions. 
BYOL\cite{DBLP:conf/nips/GrillSATRBDPGAP20} uses asymmetric architectures with predictors and stop-gradient mechanisms to prevent collapse, while DINO~\cite{DBLP:conf/iccv/CaronTMJMBJ21} leverages knowledge distillation between teacher-student networks to maintain consistency. 
Inspired by BYOL, Sang et al.~\cite{DBLP:conf/icassp/SangLLAW22} propose a self-supervised SV framework that focuses on the similarity between the latent representations of positive data pairs.  
Zhang et al. \cite{DBLP:journals/jstsp/ZhangY22} improve SV performance by applying different augmentation strategies to DINO. Chen et al.~\cite{DBLP:conf/icassp/Chen25} propose the Self-Distillation Prototypes Network (SDPN) to capture the relationship between representations of utterances from different speakers by integrating learnable prototypes into a self-distillation framework. SDPN achieves excellent performance with low computational resources, comparing favorably to many DINO-based methods.

Dimension-contrastive methods~\cite{DBLP:conf/icml/ZbontarJMLD21, DBLP:conf/iclr/BardesPL22, DBLP:conf/iclr/GarridoCBNL23} introduce a novel paradigm by optimizing cross-dimension relationships rather than instance discrimination. Barlow Twins\cite{DBLP:conf/icml/ZbontarJMLD21} minimizes correlation between dimensions to enforce independence,
while VICReg\cite{DBLP:conf/iclr/BardesPL22} learns invariance to different views, avoiding collapse of the representations with a variance preservation term, and maximizing the information content of the representation with a covariance regularization term. Building on these insights, Garrido et al.~\cite{DBLP:conf/iclr/GarridoCBNL23} demonstrate the complementary nature of sample-contrastive (including both contrastive and non-contrastive paradigms) and dimension-contrastive approaches. 

Inspired by~\cite{DBLP:conf/iclr/BardesPL22, DBLP:conf/iclr/GarridoCBNL23}, we optimize SDPN by introducing two dimension regularization terms. These additions aim to reduce the risk of collapse and improve the robustness of speaker verification by minimizing the correlation between different embedding dimensions and increasing the diversity of feature dimensions. The first regularization term is off-diagonal dimension regularization, and the second is Frobenius dimension regularization. Both techniques work to decorrelate the variables of each embedding while minimizing redundancy.

Score normalization is a common post-processing step used to align scores from different trials to a common scale. It requires no additional parameters and incurs minimal computational overhead. Most self-supervised speaker verification methods~\cite{DBLP:conf/icassp/Chen25, DBLP:conf/icassp/SangLLAW22, DBLP:journals/jstsp/ZhangY22, DBLP:conf/interspeech/HeoJKKLKC23, jin2024self, DBLP:journals/spl/ZhaoLZWZ24} employ cosine similarity for scoring. However, score distributions often vary due to factors such as channel, duration, and gender differences between datasets. A fixed threshold can adversely affect overall verification performance and exacerbate the score drift phenomenon, which is particularly pronounced in self-supervised learning.

To address this issue, we integrate normalization algorithms from supervised frameworks into SDPN with dimension regularization. This approach achieves new state-of-the-art results on VoxCeleb1, with equal error rates of 1.29\%, 1.60\%, and 2.80\% for the VoxCeleb1-O, VoxCeleb1-E, and VoxCeleb1-H trials, without using any speaker labels during training.

\section{Preliminary Knowledge}

\subsection{Self-distillation prototypes network}

The self-distillation prototypes network~\cite{DBLP:conf/icassp/Chen25} integrates a teacher and a student network, both sharing the same architecture but differing in parameters. As a non-contrastive self-supervised framework, it uses the teacher network's outputs as targets to optimize the student network concurrently. Each network comprises three primary components: an encoder for extracting speaker embeddings, a multi-layer perceptron for feature transformation, and learnable prototypes to compute soft distributions between global and local views. Here, global views $\textbf{X}_g = \{\textbf{x}_{g}\}$ are long segments, while local views $\textbf{X}_l = \{\textbf{x}_{l_1}, \textbf{x}_{l_2}, \textbf{x}_{l_3}, \textbf{x}_{l_4}\}$ are short segments, both randomly derived from the same utterance.
Consider the teacher module as an example: it consists of a backbone $f$ and a projection head $h$. The speaker embedding is derived from the output of the backbone $f$. The projection head $h$ includes three fully connected layers with dimensions 2048-2048-256, followed by $L2$ normalization. The learnable prototypes $\textbf{C}$ are shared between the teacher and student networks and are used to compute the soft distributions between global and local views. The cross-entropy loss is then calculated to minimize the probability distribution as follows:
\begin{equation}
    \mathcal{L}_{CE} = \sum_{\textbf{x} \in \textbf{X}_{g}}\sum_{\substack{\textbf{x}^{\prime} \in \mathbf{X}_{l}}} H(P^{tea}(\textbf{x}) \mid P^{stu}(\textbf{x}^{\prime}))
\end{equation}
where $H(a | b)=-a*\log b$ is cross-entropy. $P^{tea}$ and $P^{stu}$ denote the output probability distributions of the teacher network and the student network.

To prevent the collapse problem, SDPN introduces a diversity regularization term. This approach assesses the pairwise similarity among embeddings ($\textbf{x}_u$ and $\textbf{x}_v$), deliberately separating the closest embeddings to enhance the diversity of speaker embeddings within a batch, thereby enriching the learning process and enhancing the model's generalization capabilities. The diversity regularization loss is calculated as follows:
\begin{equation}
\label{eq:diversity_regularization}
    \mathcal{L}_{RE} = -\frac{1}{n} \sum_{u=1}^n (\sum_{v=1}^n \log(\min_{v \neq u} || \mathbf{x}_u - \mathbf{x}_v ||))
\end{equation}
where $n$ is the batch size.

The overall training objective of SDPN is to minimize a combination of the CE loss and diversity regularization loss, weighted by the hyperparameter $\mu$.
\begin{equation}
\label{eq:overall_loss}
  \mathcal{L}_{SDPN} = \mathcal{L}_{CE} + \mu \mathcal{L}_{RE}
\end{equation}

\subsection{Score normalization in supervised SV}

The score distribution of the test set is typically unknown in advance. Researchers propose various score normalization methods~\cite{DBLP:journals/dsp/AuckenthalerCL00, DBLP:conf/interspeech/AronowitzIB05, DBLP:conf/interspeech/MatejkaNPBSC17} to minimize the discrepancy between estimated and true distributions, such as zero normalization (Z-norm)~\cite{DBLP:journals/dsp/AuckenthalerCL00}, test normalization (T-norm)~\cite{DBLP:journals/dsp/AuckenthalerCL00}, symmetric normalization (S-norm)~\cite{DBLP:conf/interspeech/AronowitzIB05}, and adaptive symmetric score normalization (AS-norm)~\cite{DBLP:conf/interspeech/MatejkaNPBSC17}. Z-norm normalizes the score distribution of the target speaker model. T-norm normalizes the score distribution of the impostors. S-norm combines the benefits of Z-norm and T-norm by averaging their normalized scores, ensuring that the score remains consistent. Furthermore, AS-norm introduces adaptive cohort selection, choosing the top $K$ scores to calculate normalization parameters, thus aligning the estimated score distribution more closely with the actual test set distribution. It involves the use of a cohort $U = \{u_i\}_{i=1}^N$ consisting of $N$ speakers. $S_e$ are obtained by scoring the enrollment utterance $e$ against all files from the cohort $U$. $S_t$ are obtained by scoring the test utterance $t$ against all files from the cohort $U$. Specifically, the AS-norm formula is given by:
\begin{equation}  
\begin{split}
    s(e,t)_{as-norm} &= \frac{1}{2} \cdot (\frac{s(e,t) - \mu(S_e(U_e^{top}))}{\sigma(S_e(U_e^{top}))} \\
    &+ \frac{s(e,t) - \mu(S_t(U_t^{top}))}{\sigma(S_t((U_t^{top})))})
\end{split}
\end{equation}
where $s(e,t)$ is the original score between the enrollment utterance $e$ and the test utterance $t$. $U_e^{top}$ and $U_t^{top}$ represent the selected top $K$ cohort members for the enrollment and test utterances, respectively. $\mu(S_e(U_e^{top}))$ and $\sigma(S_e(U_e^{top}))$ denote the mean and standard deviation of the top $K$ cohort scores $S_e$ for the enrollment utterance $e$. $\mu(S_t(U_t^{top}))$ and $\sigma(S_t(U_t^{top}))$ are the mean and standard deviation of the top $K$ cohort scores $S_t$ for the test utterance $t$.

\begin{figure*}[hbt]
  \centering
  \includegraphics[scale=0.5]{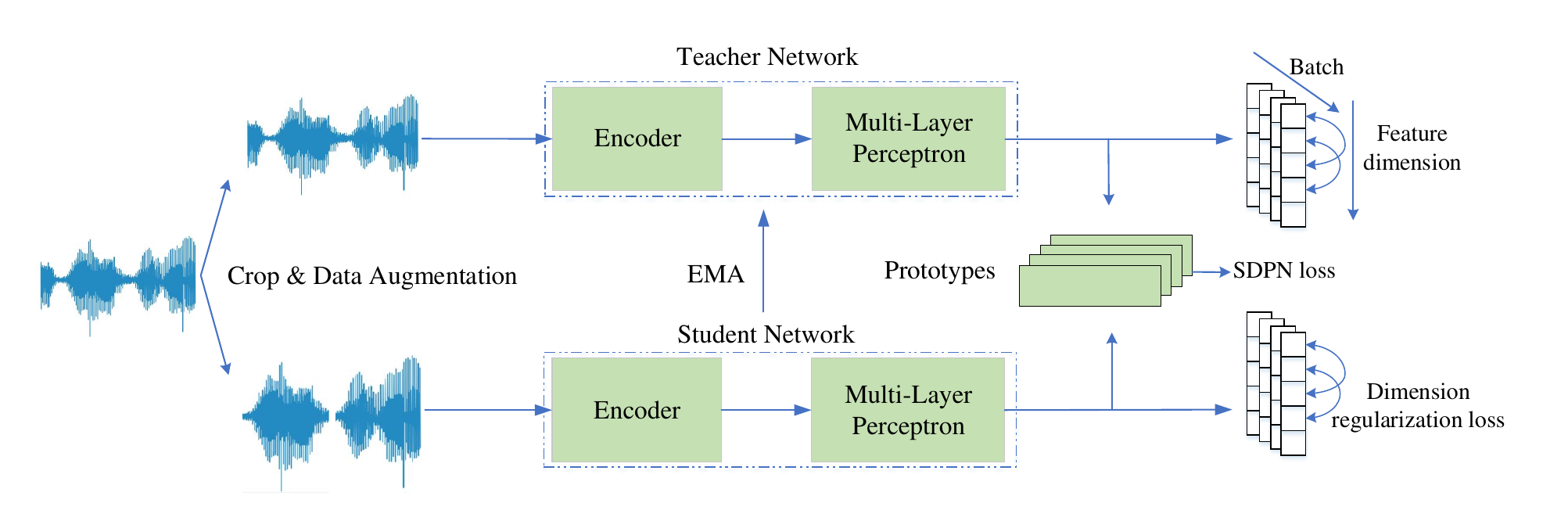}
  \vspace{-2mm}
  \caption{Overview of the SDPN framework with dimension regularization: It includes teacher and student networks with identical architectures but different parameters. The teacher network's outputs serve as targets to optimize the student network. Diversity regularization reduces the correlation between feature dimensions.}
  \label{fig:SDPN_dimension_regularization}
\end{figure*}

\section{Proposed Method}

To further alleviate the collapse problem within the SDPN framework, we introduce two dimensional regularization terms: off-diagonal dimension regularization and Frobenius dimension regularization. Additionally, we boost the SDPN with dimension regularization by incorporating various score normalization algorithms, which significantly improve the performance of self-supervised speaker verification.

\subsection{Off-diagonal dimension regularization}

The off-diagonal dimension regularization term is designed to decorrelate the variables within each embedding. By driving the covariances of each dimension in all embeddings within a batch towards zero, this approach prevents informational collapse, where variables become highly correlated. The off-diagonal dimension regularization loss is computed as follows:

\begin{equation}
  L_{ODR} = \sum_{i}^{d} \sum_{j \neq i}^{d} {C_{ij}^{tea}}^2 + \sum_{i}^{d} \sum_{j \neq i}^{d} {C_{ij}^{stu}}^2
\end{equation}
where $C$ is the covariance matrix computed from the global outputs of teacher network and student network along the batch dimension, and $d$ is matrix dimension. $C_{ij}$ is defined as Eq. \ref{eq:C_ij}:
\begin{equation}
\label{eq:C_ij}
C_{i j} = \frac{\sum_b z_{b, i} z_{b, j}}{\sqrt{\sum_b\left(z_{b, i}\right)^2} \sqrt{\sum_b\left(z_{b, j}\right)^2}}
\end{equation}
where $b$ indexes batch samples and $i,j$ index the embedding dimension. This term encourages the off-diagonal coefficients of $C_{ij}$ to be close to 0, decorrelating the different dimensions of the embeddings and preventing them from encoding similar information.

\subsection{Frobenius dimension regularization}

The Frobenius dimension regularization term minimizes the correlation between different embedding dimensions, aiming to reduce redundancy and enhance feature diversity across dimensions. It is calculated as the logarithm of the squared Frobenius norm of the normalized covariance embedding matrices, denoted as $C$. The Frobenius norm of $C$ is defined as follows:
\begin{equation}
  ||C||_{Frob} = \sqrt{\sum_{i}^{d} \sum_{j}^{d} C_{ij}^2}
\end{equation}
\begin{equation}
L_{FDR} = log(||C||_{Frob}^{tea}) + log(||C||_{Frob}^{stu})
\end{equation}
The gradient of the $L_{FDR}$ function can be formulated as:
\begin{equation}
    \frac{\partial \mathcal{L}_{FDR}}{\partial C_{ij}} = 
    \begin{cases}
        \frac{C_{ij}}{ D + \sum_{i \neq j} C_{ij}^2}, & \text{for } i \neq j \\
        0, & \text{otherwise}
    \end{cases}
\end{equation}
Different from the off-diagonal dimension regularization, this formulation exhibits two critical properties:
\begin{itemize}
    \item \textbf{Gradient magnitude modulation}: The denominator $D + \sum_{i \neq j} C_{ij}^2$ provides adaptive stabilization. When inter-dimensional correlations are weak, the gradients are bounded by the constant term $D$, preventing explosive updates. This ensures that the model's learning dynamics remain stable without excessively large gradient updates, allowing for controlled and consistent learning.
    \item \textbf{Implicit annealing effect}: As training progresses and inter-dimensional correlations diminish, the denominator undergoes a transition from being dominated by $D$ to being dominated by the correlations. This natural decay of the learning rate facilitates a more refined optimization process. 
\end{itemize}

\subsection{SDPN with dimension regularization}
The overview of SDPN framework with dimension regularization is depicted in Fig.~\ref{fig:SDPN_dimension_regularization}. 
The speaker embedding network is jointly trained with the $L_{SDPN}$ and $L_{ DR}$ ($L_{ODR}$ or $L_{FDR}$). The overall loss is calculated as Eq.~\ref{eq:overall_loss}, the hyperparameter $\lambda$ control the balance of losses.
\begin{equation}
\label{eq:overall_loss}
  \mathcal{L} = \mathcal{L}_{SDPN} + \lambda \mathcal{L}_{DR}
\end{equation}

\subsection{Score normalization in self-supervised framework}

In the field of self-supervised speaker verification, limited research has addressed score normalization to tackle inconsistent score distribution, despite the more pronounced issue of score drifting compared to fully supervised scenarios.

Among the various self-supervised frameworks available, we choose the SDPN framework with dimension regularization for its outstanding performance to evaluate the effectiveness of several score normalization techniques. 
Z-norm and T-norm are fundamental approaches that standardize scores based on global statistics, including the mean and standard deviation. AS-norm merges the benefits of Z-norm and T-norm. Its normalization parameters are calculated using the mean and standard deviation of the highest scores. This methodology aids in diminishing the discrepancy between the estimated score distribution and the actual distribution within the test set, thereby mitigating the loss of distributional information when employing a fixed quantity of score statistics as parameters.

\section{Experiments and analysis}

\vspace{-0.1cm}
\begin{table*}[thb]
    \caption{Results on VoxCeleb1-O, VoxCeleb1-E, and VoxCeleb1-H datasets. DINO* and SDPN* refers to our replication of the DINO and SDPN framework respectively. SDPN w/ off-diagonal denotes adding the off-diagonal dimension regularization to SDPN. SDPN w/ FroNorm denotes adding the Frobenius dimension regularization to SDPN. The best results for each test set are in bold.}
    \vspace{-2mm}
    \label{tab:main_results}
    \centering
    \setlength{\tabcolsep}{4pt}
    \renewcommand{\arraystretch}{0.9}
    \begin{tabular}{c c c c c c c c}
    \toprule
    \multirow{2}{*}{\textbf{Architecture}} & \multirow{2}{*}{\textbf{Score Normalization}} & \multicolumn{2}{c}{\textbf{VoxCeleb1-O}} & \multicolumn{2}{c}{\textbf{VoxCeleb1-E}} & \multicolumn{2}{c}{\textbf{VoxCeleb1-H}} \\
    \cmidrule(lr){3-4} \cmidrule(lr){5-6} \cmidrule(lr){7-8}
    & & \textbf{EER (\%)}$\downarrow$ & \textbf{minDCF}$\downarrow$ & \textbf{EER (\%)}$\downarrow$ & \textbf{minDCF}$\downarrow$ & \textbf{EER (\%)}$\downarrow$ & \textbf{minDCF}$\downarrow$ \\
    \midrule
    DINO* & Cosine & 2.65 & 0.202 & 2.74 & 0.188 & 5.02 & 0.304 \\
    SDPN* & Cosine & 1.80 & 0.139 & 1.99 & 0.131 & 3.62 & 0.219 \\
    \midrule
    \multirow{4}{*}{SDPN w/ off-diagonal} 
    & Cosine & 1.69 & 0.128 & 1.89 & 0.123 & 3.43 & 0.208\\
    & Z-norm & 1.48 & 0.119 & 1.77 & 0.113 & 3.01 & 0.194 \\
    & T-norm & 1.52 & 0.115 & 1.76 & 0.116 & 3.04 & 0.192 \\
    & AS-norm & 1.39 & 0.102 & 1.71 & 0.103 & 2.87 & 0.176 \\
    \midrule
    \multirow{4}{*}{SDPN w/ FroNorm} 
    & Cosine & 1.63 & 0.124 & 1.86 & 0.121 & 3.38 & 0.203 \\
    & Z-norm & 1.37 & 0.111 & 1.68 & 0.103 & 2.95 & 0.188 \\
    & T-norm & 1.37 & 0.105 & 1.66 & 0.105 &2.93 & 0.187 \\
    & AS-norm & \textbf{1.29} & \textbf{0.096} & \textbf{1.60} & \textbf{0.094} & \textbf{2.80} &\textbf{0.169} \\
    \bottomrule
    \end{tabular}
\end{table*}

\subsection{Experimental settings}
\label{ssec:subhead}

\subsubsection{Datasets and evaluation metrics}
\label{sssec:subsubhead}

To evaluate the effectiveness of the proposed method, we conduct experiments using the VoxCeleb datasets. The development portion of VoxCeleb2 \cite{DBLP:conf/interspeech/ChungNZ18}, consisting of 1,092,009 utterances from 5,994 speakers, is utilized for training. The performance of all systems is assessed on the test set of VoxCeleb1 \cite{DBLP:conf/interspeech/NagraniCZ17}. No speaker labels are used during training in any of the experiments. The results are presented using two metrics: the equal error rate (EER) and the minimum of the normalized detection cost function (MinDCF), with the parameters set to $P_{target}$ = 0.05 and $C_{fa} = C_{miss}$ = 1.

\subsubsection{Input features}
\label{sssec:subsubhead}
For each utterance, we employ a multi-crop strategy in SDPN training, utilizing 4-second segments as global views and 2-second segments as local views. The acoustic features in the experiments are 80-dimensional Filter Bank (FBank) with 25ms windows and a 10ms shift. Speech activity detection (SAD) is not applied, as the training data predominantly comprises continuous speech. Mean and variance normalization are conducted using instance normalization on the FBank features. WavAugment and SpecAugment are used in training process.

\vspace{-0.1cm}
\begin{table}[t]
    \caption{Comparison of the results of SDPN with FroNorm and AS-norm against those of recent SSL models on the VoxCeleb-O.}
    \vspace{-2mm}
    \label{tab:compare_ssl}
    \centering
    \setlength{\tabcolsep}{1pt}
    \begin{tabular}{llc}
    \toprule
    \textbf{Model} & \textbf{Extractor} & \textbf{EER(\%)}$\downarrow$ \\
    \midrule
    SSReg~\cite{DBLP:conf/icassp/SangLLAW22} & Fast ResNet34 & 6.99 \\
    MoCo-DSVAE~\cite{DBLP:journals/taslp/TuMC24} & ECAPA-TDNN & 6.29 \\
    Mixup-Aug~\cite{DBLP:conf/icassp/ZhangJCLHS22} & Fast ResNet34 & 5.84 \\
    DINO + CL \cite{DBLP:conf/interspeech/HeoJKKLKC23} & ECAPA-TDNN & 4.47 \\
    DINO~\cite{DBLP:journals/jstsp/ZhangY22} & ECAPA-TDNN & 3.30 \\
    {MeMo-CTES}~\cite{jin2024self} & ECAPA-TDNN & 3.10 \\
    {PDC-DINO}~\cite{DBLP:journals/spl/ZhaoLZWZ24} & ECAPA-TDNN & 3.05 \\
    C3-DINO~\cite{DBLP:journals/jstsp/ZhangY22} & ECAPA-TDNN & 2.50 \\
    SDPN~\cite{DBLP:conf/icassp/Chen25} & ECAPA-TDNN & 1.80 \\
    \midrule
    \textbf{SDPN w/ FroNorm} & \textbf{ECAPA-TDNN} & \textbf{1.63} \\
    \textbf{SDPN w/ FroNorm + AS-norm} & \textbf{ECAPA-TDNN} & \textbf{1.29} \\
    \bottomrule
    \end{tabular}
\end{table}

\vspace{-0.1cm}
\begin{figure}[t]
  \centering
  \includegraphics[scale=0.32]{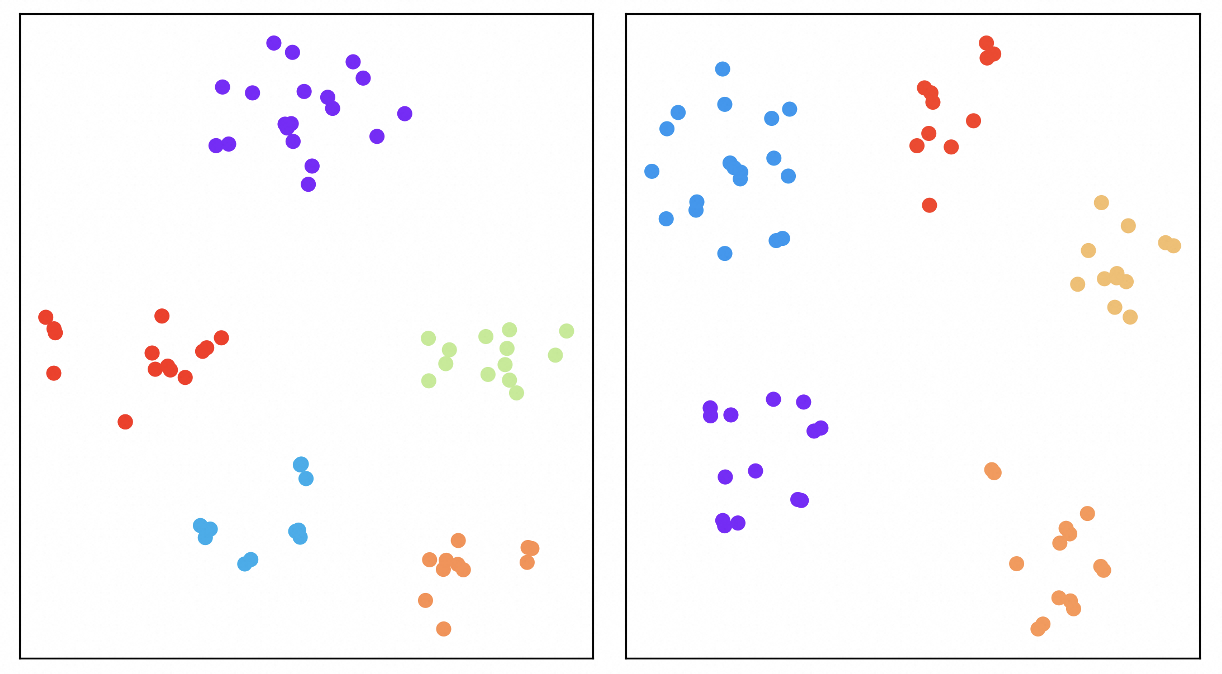}
  \vspace{-2mm}
  \caption{The t-SNE visualization presents extracted embeddings for five speakers, each represented by a different color. The left figure shows embeddings from SDPN, while the right illustrates those from SDPN with dimension regularization. The embeddings with dimension regularization demonstrate improved separation, indicating enhanced discriminability.}
  \label{fig:visualization}
\end{figure}

\subsection{Model configurations and implementation details}
\label{ssec:subhead}
We exploit the ECAPA-TDNN with attentive statistical pooling as the encoder $f$, followed by a 512-d FC layer. The projection head $h$ consists of four FC layers with hidden size of 3072-3072-1024. We train the model 160 epochs using the stochastic gradient descent (SGD) optimizer with momentum of 0.9, on 8 NVIDIA A800 GPUs. The learning rate scheduling starts with 10 warm-up epochs with a linear increase from 0 to 0.5, followed by a cosine decay with a final learning rate of 1e-5.
\subsection{Results and analysis}
\label{ssec:subhead}

The experimental results for the VoxCeleb datasets are outlined in Table~\ref{tab:main_results}. In the field of self-supervised speaker verification, DINO has been widely adopted, while SDPN represents the latest state-of-the-art model. Both have been reproduced to serve as baseline. Comparing row 1 and 2 shows that SDPN outperforms DINO substantially and consistently across all test sets. The comparison between the results of row 2 and row 3 demonstrates the effectiveness of incorporating off-diagonal dimension regularization in SDPN, achieving EERs of 1.69\%, 1.89\%, and 3.43\%, respectively. Meanwhile, the results from row 2 and row 7 show that adding FroNorm dimension regularization in SDPN further enhances verification performance, achieving relative improvements in EER by 9.44\%, 6.53\%, and 6.63\%, and in MinDCF by 10.8\%, 7.63\%, and 7.31\%, respectively. 

The use of score normalization further advances the performance of each system. The addition of Z-norm, T-norm, and AS-norm results in significant performance improvements. Due to the lack of labels in self-supervised systems, score shifts become more pronounced compared to fully supervised speaker verification systems. SDPN with \textbf{Frobenius dimension regularization and AS-norm} decreases the EER from 1.80\%, 1.99\%, and 3.62\% to \textbf{1.29\%}, \textbf{1.60\%}, and \textbf{2.80\%}, respectively, with relative improvements of \textbf{28.3\%}, \textbf{19.6\%}, and \textbf{22.6\%} on the three test sets. This represents a substantial performance enhancement, marking a further narrowing of the gap between self-supervised and fully supervised performance. 

Additionally, we utilize t-distributed Stochastic Neighbor Embedding (t-SNE)\cite{van2008visualizing} to visually assess the disentanglement performance of speaker embeddings obtained from SDPN and SDPN with Frobenius dimension regularization, as shown in Fig.~\ref{fig:visualization}. The embeddings extracted using SDPN with FroNorm dimension regularization demonstrate superior clustering capabilities, indicating that the speaker embeddings are more discriminative. Furthermore, we compare our method with recently proposed self-supervised learning architectures, including those from \cite{DBLP:conf/icassp/SangLLAW22, DBLP:journals/taslp/TuMC24, DBLP:conf/icassp/ZhangJCLHS22, DBLP:conf/interspeech/HeoJKKLKC23, DBLP:journals/jstsp/ZhangY22, jin2024self, DBLP:conf/icassp/Chen25, DBLP:journals/spl/ZhaoLZWZ24}, as detailed in Table~\ref{tab:compare_ssl}.

\section{Conclusion}

Our work narrows the performance gap between self-supervision and full supervision by introducing self-distillation prototypes network that incorporates dimension regularization and adaptive score normalization. Dimension regularization mitigates the collapse problem by enhancing feature diversity and reducing redundancy. Score normalization effectively tackles the critical issue of score drifting in self-supervised learning. Together, dimension regularization and score normalization enhance the accuracy of speaker verification.

\bibliographystyle{IEEEtran}
\bibliography{mybib}

\end{document}